\documentclass{aa}
\usepackage[utf8]{inputenc}
\usepackage[varg]{txfonts}
\usepackage{physics}
\usepackage{gensymb}
\usepackage{nameref}
\usepackage{listings}
\usepackage{balance}
\begin{document}
\title{Deceleration of kicked objects due to the Galactic potential}
\titlerunning{Deceleration of kicked objects due to the Galactic potential}

\author{P. Disberg \inst{1}
 \and N. Gaspari\inst{1}
 \and A. J. Levan\inst{1,2}
 }

\institute{Department of Astrophysics/IMAPP, Radboud University, P.O. Box 9010, 6500 GL Nijmegen, The Netherlands\\
e-mail: paul.disberg@gmail.com
\and
Department of Physics, University of Warwick, Coventry CV4 7AL, UK}

\date{\today}

\abstract{Various stellar objects experience a velocity kick at some point in their evolution. These include neutron stars and black holes at their birth or binary systems when one of the two components goes supernova. For most of these objects, the magnitude of the kick and its impact on the object dynamics remains a topic of debate.}
{We investigate how kicks alter the velocity distribution of objects born in the Milky Way disc, both immediately after the kick and at later times, and whether these kicks are encoded in the observed population of Galactic neutron stars.}
{We simulate the Galactic trajectories of point masses on circular orbits in the disc after being perturbed by an isotropic kick, with a Maxwellian distribution of magnitudes with $\sigma=265$ km/s. Then, we simulate the motion of these point masses for $200$ Myr. These trajectories are then evaluated, either for the Milky Way population as a whole or for those passing within two kiloparsecs of the Sun, to get the time evolution of the velocities.}
{During the first $20$ Myr, the bulk velocity of kicked objects becomes temporarily aligned to the cylindrical radius, implying an anisotropy in the velocity orientations. Beyond this age, the velocity distribution shifts toward lower values and settles to a median of $\sim200$ km/s. Around the Sun, the distribution also loses its upper tail, primarily due to unbound objects escaping the Galaxy. We compare this to the velocities of Galactic pulsars and find that pulsars show a similar evolution with characteristic age.}
{The shift of the velocity distribution is due to bound objects spending most of their orbits at larger radii after the kick. They are, therefore, decelerated by the Galactic potential. We find the same deceleration to be predicted for nearby objects and the total population and conclude it is also observed in Galactic pulsars. Because of this effect, the (scalar) speeds of old neutron stars provide little information about their kicks at birth.}
\maketitle

\section{Introduction}
\label{sec1}
Many astrophysical objects experience a velocity kick during their evolution. The most notable case is that of neutron stars, and in particular of pulsars, which have space velocities significantly higher than their progenitors and thus must have received an additional velocity - a natal kick - at birth \citep{Gunn_1970,Lyne_1994,Cordes_1998}. Other instances of likely kicks include stellar-mass black holes \citep{Jonker_2004,Fragos_2009,Repetto_2012,Mandel_2016,Andrews_2022}, 
and binary systems, which receive a systemic kick due to mass loss after one of the two components goes supernova \citep{Blaauw_1961,Hills_1983,Brandt_1995,VandenHeuvel_2000,Willems_2004,Zhang_2013,Atri_2019,Gandhi_2019,Fortin_2022,Zhao_2023}.\\
\indent There remain several open questions regarding these kicks, the most pressing of which, for most systems is their magnitudes. For pulsars, some studies found that they were best described with a single-peaked distribution with relatively high values \citep[e.g.,][]{Hobbs_2005,Faucher_2006}, while more recent studies found a double-peaked distribution with a second significant peak at lower values \citep{Arzoumanian_2002,Verbunt_2017,Igoshev_2020}. The distribution of kick magnitudes, in turn, has implications not only for our understanding of the nature of the kicks, for example whether they are due to binary disruption or are intrinsic to an asymmetric supernova \citep{Dewey_1987,VanParadijs_1995,Iben_1996,Fryer_1997,VandenHeuvel_1997}, but also for the progenitors evolution \citep[e.g.,][]{Kalogera_1996,Podsiadlowski_2004,VandenHeuvel_2007,Beniamini_2016}, the supernova mechanism \citep[e.g.,][]{Janka_2012,Janka_2013,Wongwathanarat_2013,Coleman_2022,Burrows_2024}, the merger rates of compact object binaries \citep[e.g.,][]{Dominik_2012,Giacobbo_2018,Kruckow_2018,Vigna_2018,Iorio_2023} and the locations of the transients they produce \citep[e.g.,][]{Fryer_1999,Perna_2002,Voss_2003,Belczynski_2006,Church_2011,Jiang_2020,Zevin_2020,Mandhai_2022}.\\
\indent Kicks kinematically decouple a population of objects from their progenitors, thus determining their present-day locations and velocities. In the context of Galactic pulsars, several authors have modelled their Galactic trajectories in order to constrain the phase-space distribution and compare to observations. This method have been used to check whether pulsar could have been a viable gamma-ray burst progenitor \citep{Hartmann_1990,Paczynski_1990}, to constrain the evolution of pulsars spins and magnetic fields \citep{Bhattacharya_1992,Hartman_1997}, and to study the population of isolated old neutron stars which might become observable due to accretion of the interstellar medium \citep{Blaes_1991,Blaes_1993,Zane_1995,Popov_2000,Treves_2000}. More recent works modelling the positions and velocities of pulsars across the Galaxy include those of \citet{Kiel_2009}, \citet{Ofek_2009}, \citet{Sartore_2010}, and \cite{Sweeney_2022}.\\
\indent The distribution of velocities in a population of kicked objects evolves with time, but most importantly, it also varies with the Galactic location where it is probed. That is to say, the velocity distribution we might be able to observe could be different from that of the total population. \citet{Lyne_1994}, for instance, observe that young pulsars have higher velocities than older ones, but do not believe this is caused by an actual dynamical evolution. Instead, they link this to the work of \citet{Cordes_1986}, who argues that high-velocity pulsars rapidly move away from the Solar neighbourhood, reducing their likelihood of detection, and resulting in an observational bias towards lower velocities for older pulsars. \citet{Hansen_1997} investigate this argument, using a Monte Carlo simulation of masses moving in a Galactic potential and analysing the Galactocentric velocities of the masses which get close to the Solar neighbourhood. They find that: (1) at first, these masses have a high velocity, (2) after approximately $10$ Myr, the radial velocity of the set increases, due to masses born close to the Galactic centre arriving at the Solar system, and (3) after approximately $100$ Myr, only the low-velocity tail of the distribution is left.\\
\indent In this paper, we are interested in tracking the apparent velocity shifts first observed in pulsars by \citet{Lyne_1994}, and understanding their origin. We aim to investigate how kicks affect the velocity of astrophysical objects, and how these effects can be detected in the Solar neighbourhood at times which may be long after the kick was imparted to the object. In order to examine the velocity evolution of kicked objects, we use a method similar to the one used by \citet{Hansen_1997}, i.e., a Monte Carlo simulation of point masses moving through the Galactic potential \citep[also, cf.][and references therein]{Sartore_2010}.\\
\indent This paper is structured as follows. In Sect.\ \ref{sec2}, we describe our model and simulation. Then, in Sect.\ \ref{sec3}, we give our main results, which show that the deceleration of kicked objects due to the Galactic potential is dominant in shaping the final velocity distribution. In Sect.\ \ref{sec4}, we explain these results by using grids of simulations which exemplify the dynamics. Moreover, in Sect.\ \ref{sec5}, we compare our results 2D and (estimated) 3D velocities of pulsars. Finally, in Sect.\ \ref{sec6}, we summarise our findings and their implications.
\section{Simulation}\label{sec2}
First, we describe our model and simulation. In particular, we describe our Milky Way (MW) model, which is used to seed the point masses in our Monte Carlo simulation (Sect.\ \ref{sec2.1}). Then, we elaborate on how we initialise and evaluate the trajectories of the objects (Sect.\ \ref{sec2.2}).
\subsection{Milky Way model}
\label{sec2.1}
In order to describe the initial positions of the point masses, we use the MW model of \citet{Gaspari_2024}, which is based on the model of \citet{Chrimes_2021}, and entails a synthetic MW image that we use to sample the initial positions of the simulated objects. The MW model consists of three parts: an exponential disc, a bar-like bulge, and spiral arms. Here, we give a brief overview of these components, while more detailed descriptions are given by \cite{Gaspari_2024} and \citet{Chrimes_2021}.\\
\indent The bulge is modelled with the triaxial boxy Gaussian distribution of \citet{Grady_2020}, where the bar makes a $27\degree$ angle with the line connecting the Solar system to the Galactic centre, in the $xy$-plane \citep{Wegg_2013}. For the spiral arms, we use the data from \citet{Chrimes_2021}, who use the method of \citet{Reid_2016,Reid_2019} to map the spiral arms, using water and methanol masers. In our 3D MW model, we set the 2D spiral arms of \citet{Chrimes_2021} at $z=0$. Also, since their arm-profile only covers the lower half of the MW, we rotate them for the upper half. For the disc, we use an exponential density profile:
\begin{equation}
    \label{eq1}
    \rho_{\text{disc}}\left(x,y,z\right)\propto\exp\left(-\dfrac{\sqrt{x^2+y^2}}{R_d}\right)\exp\left(-\dfrac{|z|}{z_d}\right)\quad,
\end{equation}
where we choose $R_d=2.6$ kpc and $z_d=0.3$ kpc \citep{Hawthorn_2016}. The density profiles of the arms, bulge, and disc are used to create a synthetic image of the MW, with a scale of $250$ pc per pixel, and create the synthetic MW image by showing the densities on a logarithmic scale (effectively converting luminosity to magnitude).\\
\begin{figure}
    \resizebox{\hsize}{!}{\includegraphics{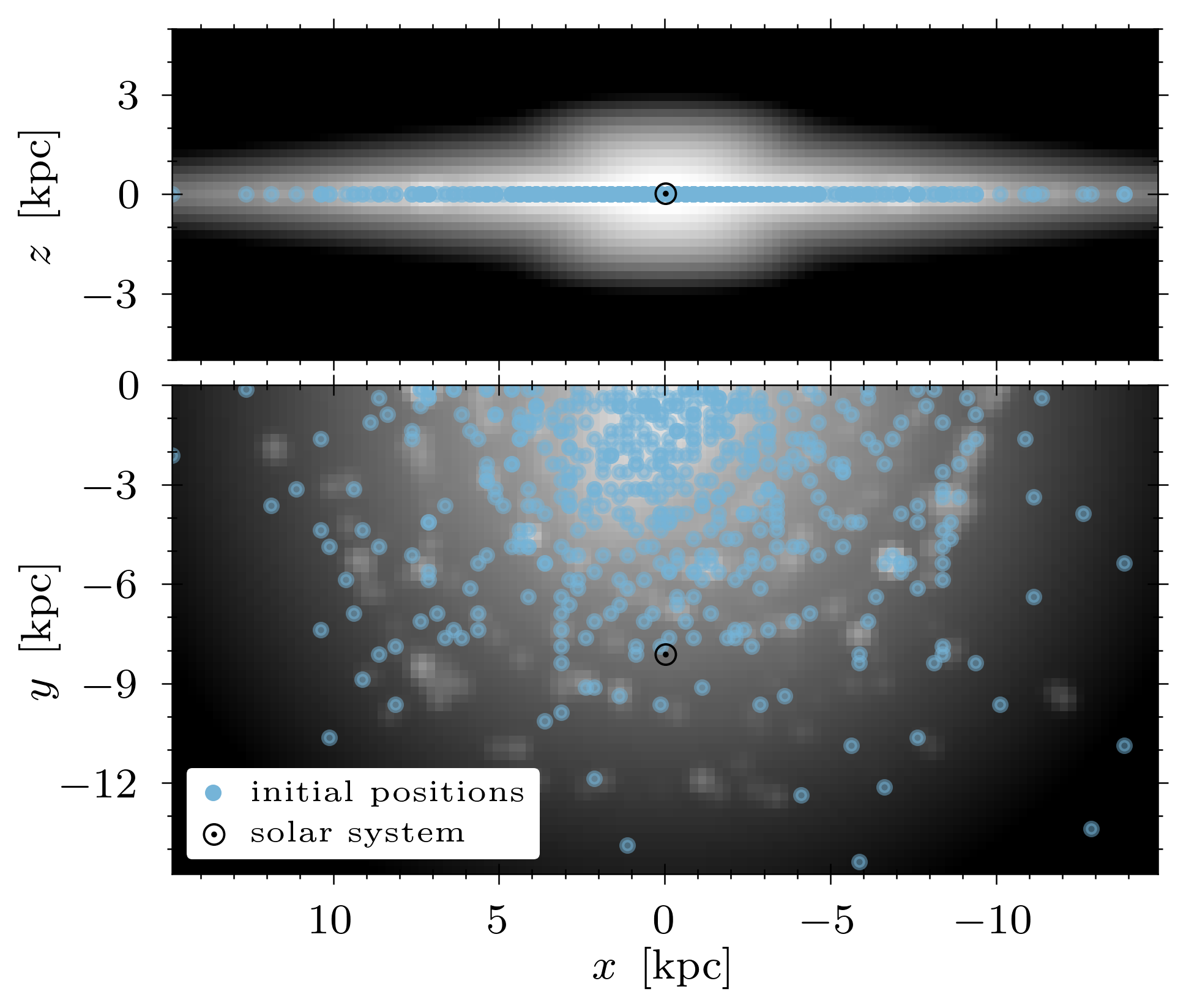}}
    \caption{Initial positions of point masses in our simulation (blue dots), plotted over the synthetic MW image, based on the works of \citet{Gaspari_2024} and \citet{Chrimes_2021}, as described in the text. The figure contains $500$ of the $10^4$ simulated point masses. The sun-symbol represents the Solar system, at $R_{\sun}=8.122$ kpc and $z_{\sun}=20.8$ pc.}
    \label{fig1}
\end{figure}
\indent The three components in this image are weighted based on their luminosity. We use the $I$-band luminosities of \citet{Flynn_2006}, who estimate a luminosity of $1\cdot10^{10}L_{\sun}$ for the bulge, and a luminosity of $3\cdot10^{10}L_{\sun}$ for the combination of spiral arms and exponential disc. Also, we choose an arm-strength of $0.15$ \citep{Gaspari_2024}, meaning the arms get a luminosity of $0.45\cdot10^{10}L_{\sun}$ and the disc gets a luminosity of $2.55\cdot10^{10}L_{\sun}$. The density profiles are normalised and multiplied by these luminosity. Moreover, we add a Gaussian blur with a FWHM of $2$ pixels \citep[similar to][]{Chrimes_2021}, in order to slightly broaden the arms and create a more realistic model.\\
\begin{figure*}
    \centering
    \includegraphics[width=18cm]{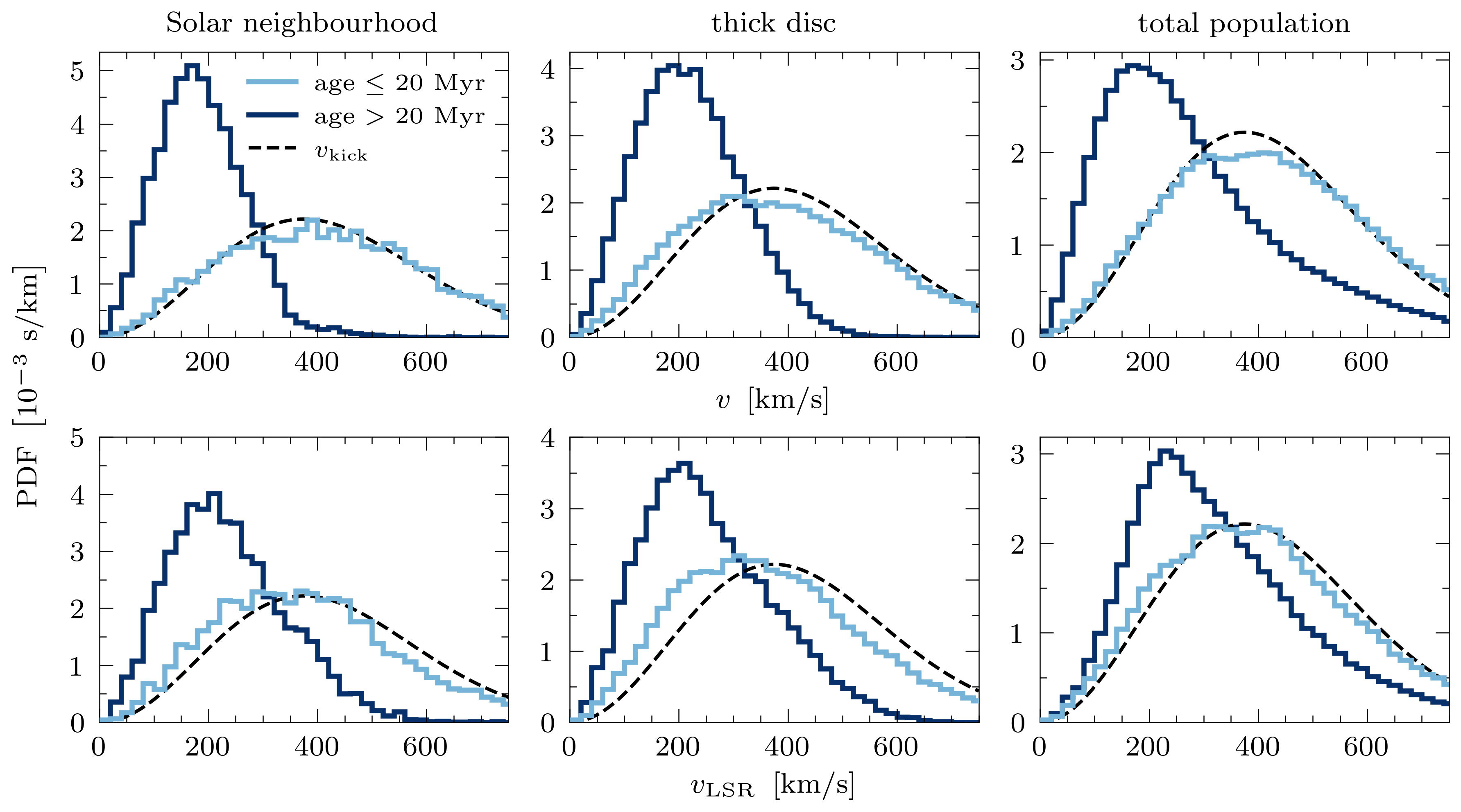}
    \caption{Distributions of Galactocentric velocities ($v=|\vec{v}|$, top row) and peculiar velocities in the Local Standard of Rest of the point mass ($v_{\text{LSR}}$, bottom row) for the simulated point masses. For the first $20$ Myr, we evaluate the velocities every $1$ Myr, and show the total, normalised distribution, in histograms with a bin-width of $20$ km/s (light blue curves). After $20$ Myr, we evaluate the velocities every $10$ Myr and also show the total, normalised distribution, in similar histograms (dark blue curves). The dashes, black curves show the kick distribution \citep[i.e., a Maxwellian with $\sigma=265$ km/s, as used by][]{Hobbs_2005}. The three panels show the evaluated velocities for the three populations: the Solar neighbourhood (left), the thick disc (centre), and the total population (right). These populations have sizes of $24817$, $144496$, and $210000$ for ages below $20$ Myr, and $7079$, $54320$, and $180000$ for ages above $20$ Myr, respectively.}
    \label{fig2}
\end{figure*}
\indent In order to seed the initial positions of the point masses in our simulation, we use this MW model. We seed our point masses in the thin disc, since kicks often occur in the context of massive --and therefore young-- stars, and this is where the population of young, massive stars is situated \citep[e.g.,][]{Feltzing_2008}. We therefore set the initial positions of our point masses at $z=0$, because the scale height of the thin disc is approximately the size of one pixel. Furthermore, we sample the initial positions in the $xy$-plane from a combination of the spiral arms and the exponential disc (Eq.\ \ref{eq1}, where $z=0$). Figure\ \ref{fig1} shows the synthetic MW image and the sampled initial positions. Our simulation consists of $10^4$ point masses, of which $500$ are shown in the figure. The image contains only the lower half of the $xy$-plane, but the points are sampled in the upper half as well (which uses a rotated version of the lower half).
\subsection{Trajectories}
\label{sec2.2}
After the initial positions have been sampled, the initial velocities of the points are determined. In order to model the kick, we use the kick velocity distribution of \citet{Hobbs_2005}, who estimate the kick velocities of pulsars to be best described by a Maxwellian with $\sigma=265$ km/s. We do not argue that their kick distribution accurately describes pulsar kicks, but use their work to show how such a distribution evolves over time. Indeed, as we show in Appendix B, the final velocity distribution is not strongly dependent on the shape of the initially assumed kick distribution. For each point, we sample a kick velocity ($v_{\text{kick}}$) from the \citet{Hobbs_2005} distribution. Assuming the kick direction is isotropic, we orient the kick vector by uniformly sampling an angle $\alpha$ within the $xy$-plane, and sampling an angle $\beta$ with a probability proportional to $\cos\beta$, which is the declination relative to the $xy$-plane. This way, the kick vectors are oriented isotropically.\\
\indent We also assume the points are in a circular orbit around the Galactic centre before they experience the kick, meaning they get an additional circular velocity ($v_{\text{circ}}$), which has a magnitude determined by the gravitational potential. We use the MW potential of \citet{McMillan_2017} to determine these circular velocities.\\
\indent After combining the kick and the circular velocity, the initial velocity vector ($\vec{v}_0$), in the cylindrical coordinates of azimuth ($\phi$), Galactocentric radius ($R$), and vertical height ($z$), becomes:
\begin{equation}
    \label{eq2}
    \vec{v}_0=\begin{pmatrix}v_{\phi}\\ v_{R}\\ v_{z}\end{pmatrix}=\begin{pmatrix}v_{\text{kick}}\sin\alpha\cos\beta+v_{\text{circ}}\hfill\\ v_{\text{kick}}\cos\alpha\cos\beta\hfill\\ v_{\text{kick}}\sin\beta\hfill\end{pmatrix}\quad.
\end{equation}
Having determined the initial positions and velocities of the point masses, we evolve their trajectory in the \citet{McMillan_2017} potential using the \lstinline{GALPY}\footnote{\href{http://github.com/jobovy/galpy}{http://github.com/jobovy/galpy}} \lstinline{v.1.9.0} package \citep{Bovy_2015}. In initialising the potential, we assume that the circular velocity at the position of the Sun equals the azimuthal velocity of the Sun.\\
\begin{figure*}
    \sidecaption
    \includegraphics[width=12cm]{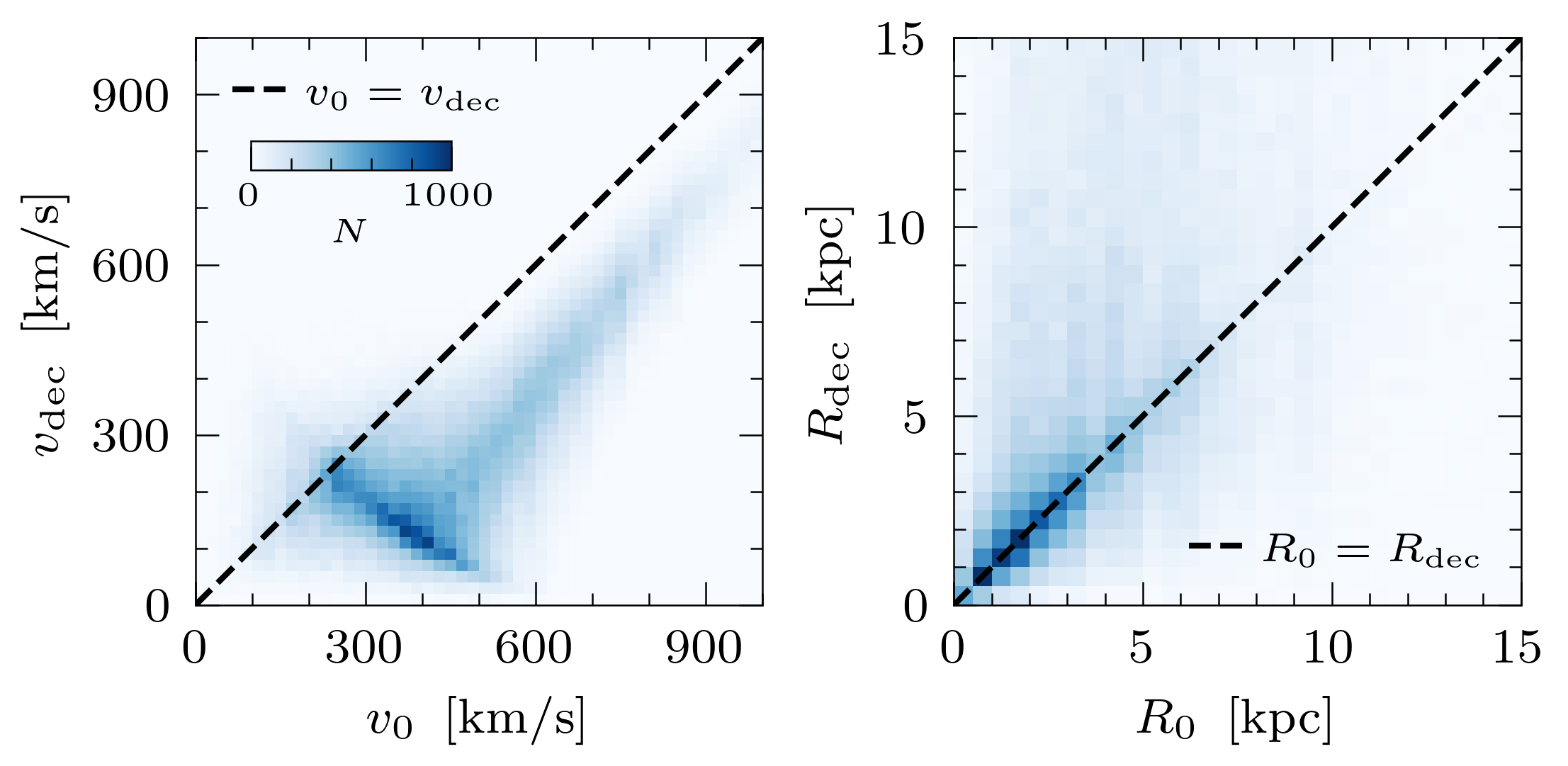}
    \caption{Initial conditions versus decelerated conditions, for the total population of our simulation. The left panel shows the initial velocities of the objects ($v_0$) versus the decelerated velocities ($v_{\text{dec}}$), which we define as any velocity for $t>20$ Myr. We evaluate the decreased velocities every $10$ Myr, and show the resulting distribution in a 2D histogram with bins of $20$ km/s. The right panel shows the initial Galactocentric radial position of the objects ($R_0$) versus the Galactocentric radial position of the objects for $t>20$ Myr ($R_{\text{dec}}$), evaluated every $10$ Myr and shown in a 2D histogram with bins of 0.5 kpc. Here, $83$\% of the objects are located at $R_{\text{dec}}>R_0$.}   
    \label{fig3}
\end{figure*}
\begin{figure}
    \resizebox{\hsize}{!}{\includegraphics{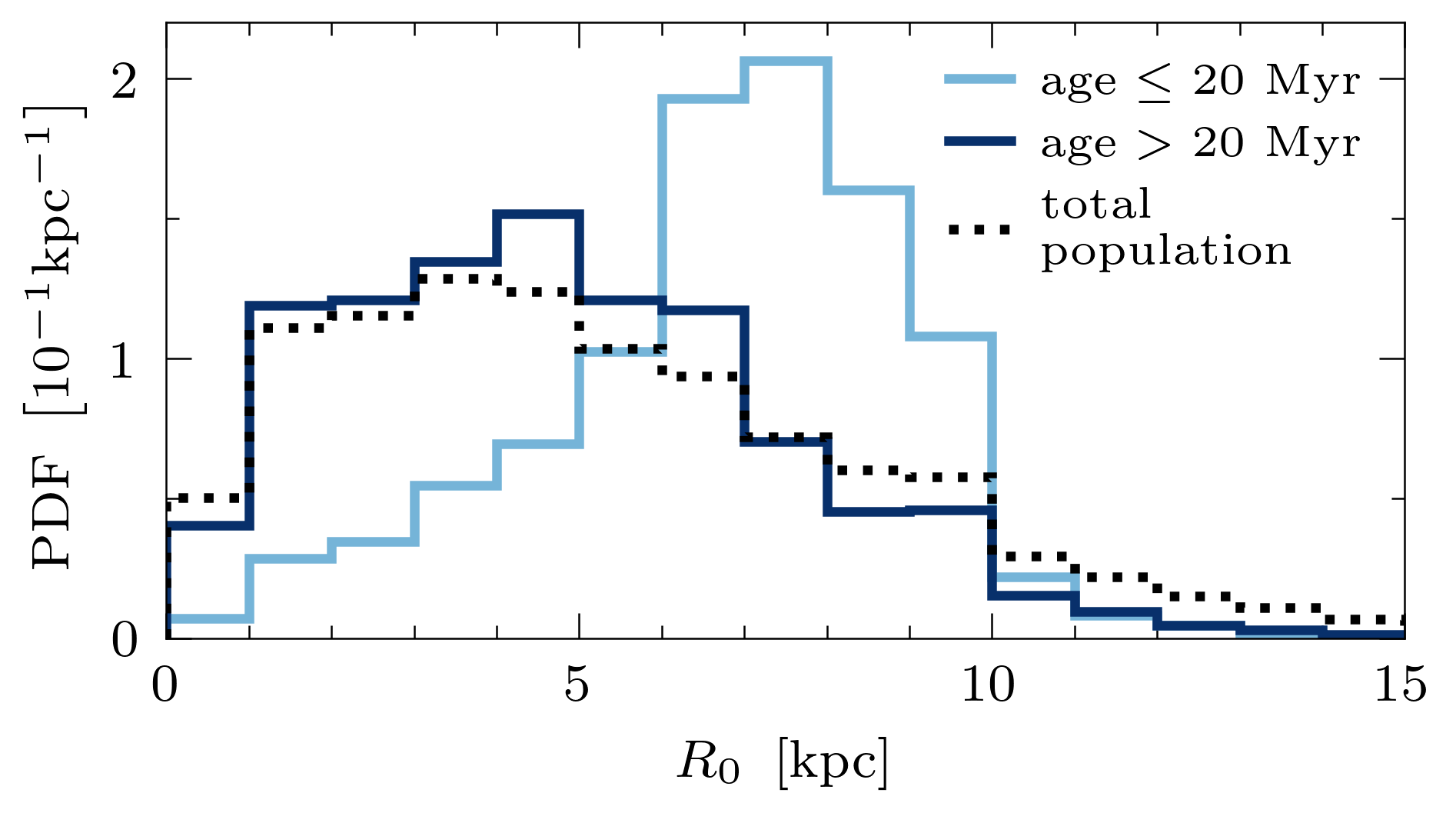}}
    \caption{Initial Galactocentric radii ($R_0$) of the objects that are in the Solar neighbourhood below $20$ Myr (light blue) and above $20$ Myr (dark blue), shown in normalised histograms with bins of $1$ kpc. The dotted curve shows the initial radii of the total population, in a similar histogram. The distribution of the total population peaks slightly above the half-light radius $R_d=2.6$ kpc (Eq. \ref{eq1}), because the spiral arms shift the initial positions to higher radii, compared to the disc.}
    \label{fig4}
\end{figure}
\indent We evolve the trajectories of the point masses for $200$ Myr, while evaluating the position and velocity of each point at several times during their trajectory. When we evaluate the trajectories at a certain time, we consider three populations in our simulation:
\begin{itemize}
    \item The Solar neighbourhood, which we define as all points in the simulation that come within a distance of $2$ kpc to a Solar orbit. We determine this by initialising $12$ orbits with the initial conditions of the Solar systems, each one azimuthally rotated by $2\pi/12$. Effectively, this creates 12 non-overlapping spheres with a radius of $2$ kpc, moving in a Sun-like orbit \citep[cf.][]{Hansen_1997}. Azimuthal rotation does not affect the results significantly, due to the MW model being almost cylindrically symmetric. We use the Solar velocity:
    \begin{equation}
        \label{eq3}
        \vec{v}_{\sun}=\begin{pmatrix}v_{\phi}\\v_{R}\\v_{z}\end{pmatrix}=\begin{pmatrix}245.6\ \text{km/s \citep{Gravity_2018}}\hfill\\-12.9\ \text{km/s \citep{Drimmel_2018}}\hfill\\7.78\ \text{km/s \citep{Reid_2004}}\hfill\end{pmatrix}\quad,
    \end{equation}
    and the Solar position (rotated for each $n^{\text{th}}$ sphere, where $0\leq n<12$):
    \begin{equation}
        \label{eq4}
        \vec{r}_{\sun}=\begin{pmatrix}\phi\\ R\\ z\end{pmatrix}=\begin{pmatrix}n\cdot2\pi/12\hfill\\8.122\ \text{kpc \citep{Gravity_2018}}\hfill\\20.8\ \text{pc \citep{Bennett_2019}}\hfill\end{pmatrix}\quad,
    \end{equation}
    as initial conditions for these Solar spheres. These values correspond to the ones implemented in \lstinline{GALPY v.1.9.0}. We co-evolve these Solar orbits and at each point in time evaluate the Galactocentric velocities of the points which are within one of these spheres.
    \item The thick disc. As a rough approximation of the thick disc, based on Fig.\ \ref{fig1}, we define this region as $R\leq15$ kpc and $|z|\leq2$ kpc.
    \item The total population, including all point masses, even the ones which get a high kick and escape the Galaxy.
\end{itemize}
Considering these three populations allows us to compare the velocities of the points which get close to Solar orbit--and therefore have a higher probability of being detected--to the velocities of the other points. This way, we can estimate the role of observational bias in context of, e.g., older pulsars appearing to have lower velocities.
\section{Deceleration}
\label{sec3}
Analysing the results from the simulation described in Sect.\ \ref{sec2}, we first investigate how the velocity distribution changes over time (Sect.\ \ref{sec3.1}), and then examine how the components of the velocity vector (i.e., $v_{\phi}$, $v_{R}$, and $v_{z}$) evolve as the points move through the Galaxy (Sect.\ \ref{sec3.2}).
\subsection{Velocity magnitude}
\label{sec3.1}
\begin{figure*}
    \centering
    \includegraphics[width=18cm]{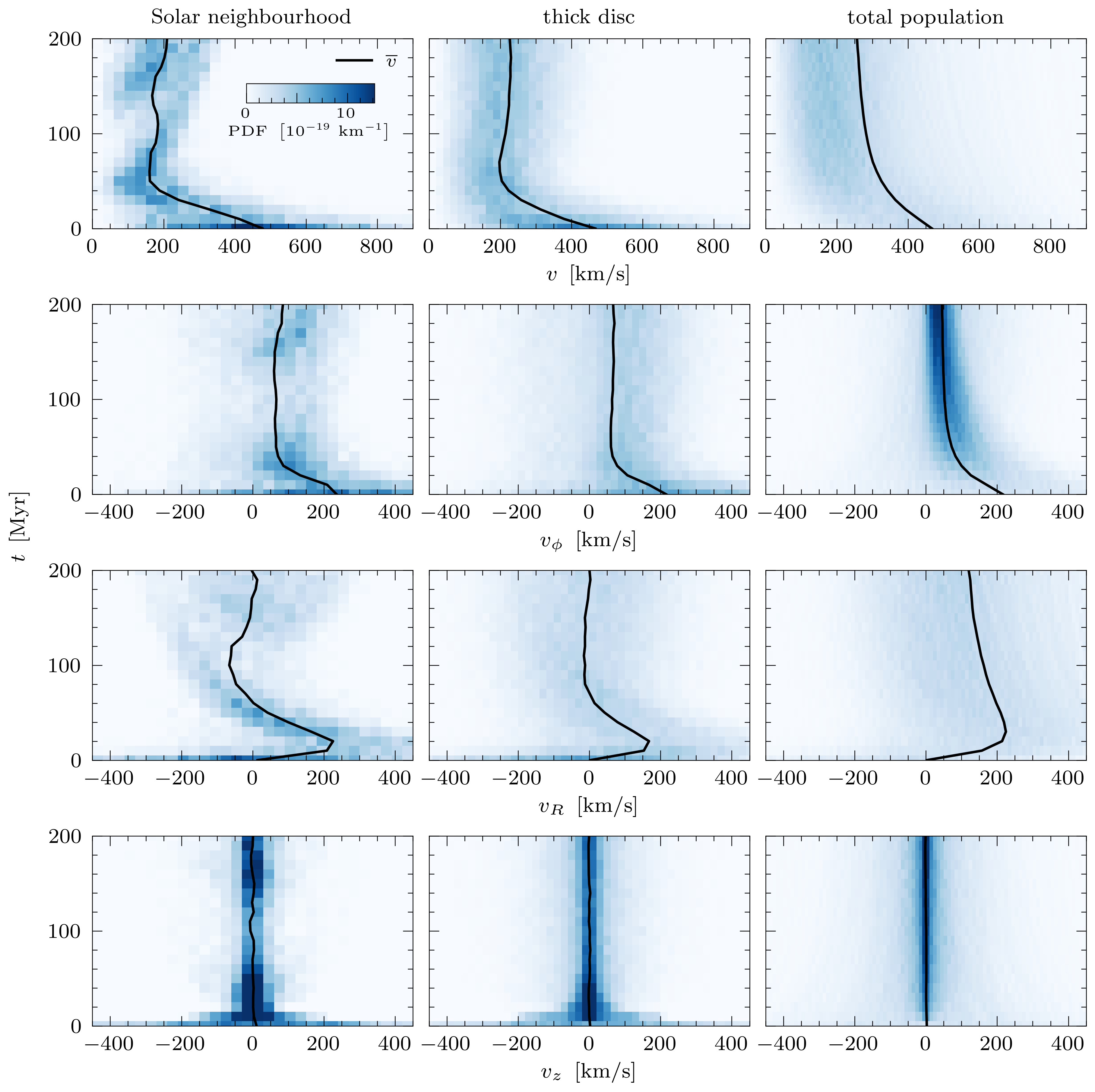}
    \caption{Evolution of the Galactocentric velocity ($v$) and the cylindrical components of the velocity vector ($v_{\phi}$, $v_{R}$, and $v_{z}$). The figure shows normalised distributions, in 2D histograms with time-bins of $10$ Myr and velocity-bins of $30$ km/s, $20$ km/s, and $10$ km/s for the Solar neighbourhood, the thick disc, and the total population, respectively. For clarity, we set the maximum values of the distributions at $4\cdot10^{-5}$ s km$^{-1}$ Myr$^{-1}$ $\approx13\cdot10^{-19}$ km$^{-1}$, even though the peaks in the last row exceed this maximum to a certain degree. The black lines show the mean velocities at each point in time.}
    \label{fig5}
\end{figure*}
First, we note that at $t=0$ Myr, the Galactocentric velocity distribution of the point masses does not precisely match the \citet{Hobbs_2005} kick distribution (in all three populations). This is caused by the circular velocity which is added to the kick velocity, i.e., the $v_{\text{circ}}$-term in Eq.\ \ref{eq2}. In other words, the velocity distribution is shifted to higher velocities at first, compared to the kick distribution, and then decreases to lower velocities. We find that if we consider all ages up to $20$ Myr, the total distribution averages out to approximately the kick distribution. For this reason, we compare the velocities below $20$ Myr to the ones above $20$ Myr.\\
\indent This comparison considers the distributions of the Galactocentric velocities, meaning they are, e.g., not corrected for the motion of the Solar system. In general, we find that the velocity distributions show a deceleration in all three populations. It is important to note here that with \textquotedblleft deceleration\textquotedblright, we are referring to the fact that after about $20$ Myr, the objects tend to have velocities lower than their initial velocity, this does not mean that the objects at these points in time are necessarily decelerating (i.e., have $dv/dt<0$). Figure\ \ref{fig2} displays the three distributions, for ages below and above $20$ Myr, and indeed shows that for ages below $20$ Myr, the velocities approximate the kick distribution, while above $20$ Myr the velocities are significantly lower. In Appendix \ref{appA}, we zoom in on the decrease in the velocity distribution, and show how it remains stable beyond $200$ Myr up to at least $1$ Gyr.\\
\indent Figure\ \ref{fig2} shows that the total distributions below $20$ Myr approximate the kick distribution. The distributions, however, start to shift to lower velocities, becoming relatively stable after $20$ to $30$ Myr. Interestingly, the decelerated velocity distributions look similar in all three populations: they show a peak around $200$ km/s. The main difference between the three populations can be found in the high-velocity tail, which is significantly smaller in the Solar neighbourhood. This is due to the objects that get a large kick, escape the Galaxy, and are therefore not found in the Solar neighbourhood but still contribute to the total population. The thick disc also shows more high-velocity objects than the Solar neighbourhood, which is probably caused by objects that get a significant kick but cannot escape the Galaxy and fall back to the Galactic centre.\\
\indent The bottom row of Fig.\ \ref{fig2} shows the distributions of $v_{\text{LSR}}$, which equals the velocity in the Local Standard of Rest (LSR) frame. That is, for an object this velocity is calculated by subtracting the local $v_{\text{circ}}$ from its velocity vector. The distributions of $v_{\text{LSR}}$ look similar to the Galactocentric velocities, and show a similar deceleration. The deceleration of the kicked objects is therefore not limited to the Galactocentric frame.\\
\indent We argue that because the distributions of the point masses above $20$ Myr decelerate and look similar in all three populations, it is difficult to explain the deceleration with an observational bias \citep[as, e.g., proposed by][]{Cordes_1986,Lyne_1994}. After all, the difference between the three populations is mainly the high-velocity tail, and if the deceleration was merely apparent, and the result of an observational bias, it should only be present in the Solar neighbourhood and absent from the total population.\\
\begin{figure*}
    \centering
    \includegraphics[width=18cm]{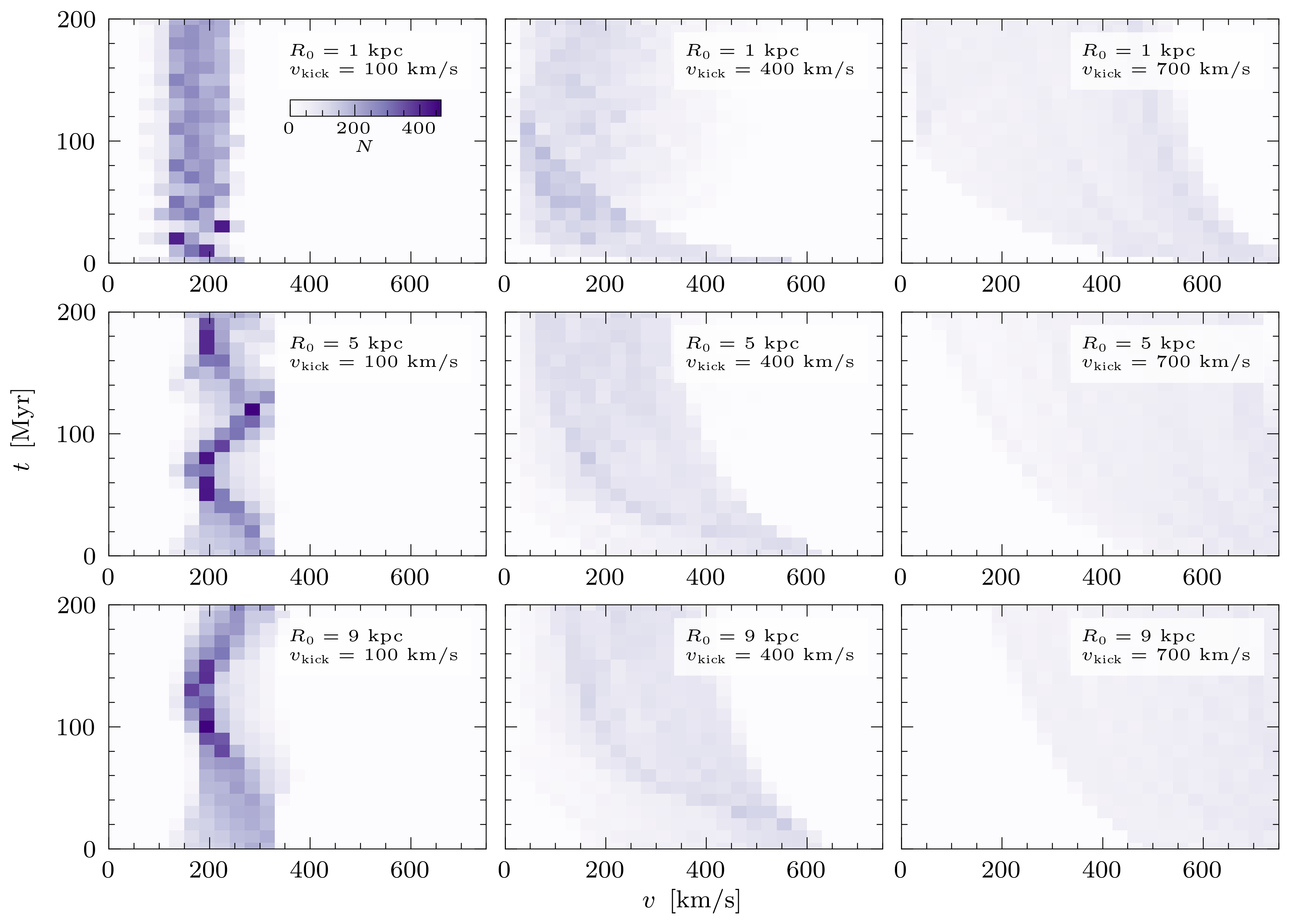}
    \caption{Grid of simulations of $10^3$ kicked objects, with a fixed initial Galactocentric radius $R_0$ (rows: $1$ kpc, $5$ kpc, and $9$ kpc) and a single value for $v_{\text{kick}}$ (columns: $100$ km/s, $400$ km/s, and $700$ km/s). The figure shows the evolution of the Galactocentric velocity ($v$) of the total population, in 2D histograms with time-bins of $10$ Myr and velocity-bins of $30$ km/s. Even though these simulations use a single value for $v_{\text{kick}}$, the velocity distributions do not start as delta functions, because of the $v_{\text{circ}}$ term in Eq.\ \ref{eq2}.}
    \label{fig6}
\end{figure*}
\indent In the left panel of Fig.\ \ref{fig3}, we show how the velocities after $20$ Myr relate to the initial velocities. The main deceleration, as can be seen in the figure, occurs for velocities below approximately $500$ km/s. Here, a higher $v_0$ results in a lower velocity after $20$ Myr. This can be explained by the fact that higher kicks disturb the initial circular orbits more significantly and cause the objects to follow new, eccentric, orbits with an apocentre at higher Galactocentric radii. In these orbits, the objects spend most of their time near the apocentre, where they have velocities lower than their initial circular velocity, due to the gravitational potential. This is why objects in the high-velocity tail (i.e., $v>500$ km/s) retain velocities closer to their initial velocity: these objects get a sufficiently large kick to escape the Galaxy and are therefore less affected by the Galactic potential if they get a higher kick. The right panel in Fig.\ \ref{fig3} shows the Galactocentric radius after $20$ Myr, compared to the initial Galactocentric radius ($R_0$), and it indeed shows that $83$\% of the objects are, after $20$ Myr, situated at radii larger than their initial $R_0$.\\
\indent We therefore argue that interaction with the gravitational potential of the Galaxy can provide a better explanation, because (1) the deceleration is prevalent in all three populations, (2) the kicks disturb the initial circular orbits of the objects, after which they obtain eccentric orbits in which they spend most of their time at higher Galactocentric radii, where they have velocities lower than their initial velocity (due to the Galactic potential), and (3) this effect does not depend on the \citet{Hobbs_2005} distribution being exactly right, since Maxwellians with a different $\sigma$ show a similar deceleration (in Appendix \ref{appB}, we show how varying the $\sigma$ of these Maxwellians gives indistinguishable decelerated distributions in the Solar neighbourhood).\\
\indent The hypothesis that the MW potential determines this deceleration is also supported by the initial Galactocentric radii of the objects which arrive in the Solar neighbourhood. Figure\ \ref{fig4} shows the $R_0$ distributions of the objects in the Solar neighbourhood, both below and above $20$ Myr. At $t=0$ Myr, all objects in the Solar neighbourhood are born at an initial Galactocentric radius where $|R_0-R_{\sun}|\leq2$, and the distribution for age $\leq20$ Myr still shows a vast majority of objects born in this region. For age $>20$ Myr, however, enough time has passed for the kicked objects to have travelled relatively large distances, because of which the initial bias towards locally born objects disappears and the distribution starts to resemble the initial radii of the total population. After $20$ Myr, therefore, the Solar neighbourhood is a representative sample of the total population, both in velocity magnitude and birth locations. In other words: after a certain amount of time, the objects in the Solar neighbourhood are mostly born significantly closer to the Galactic centre, meaning they migrated to larger radii and lost kinetic energy due to the Galactic potential.
\begin{figure*}
    \centering
    \includegraphics[width=18cm]{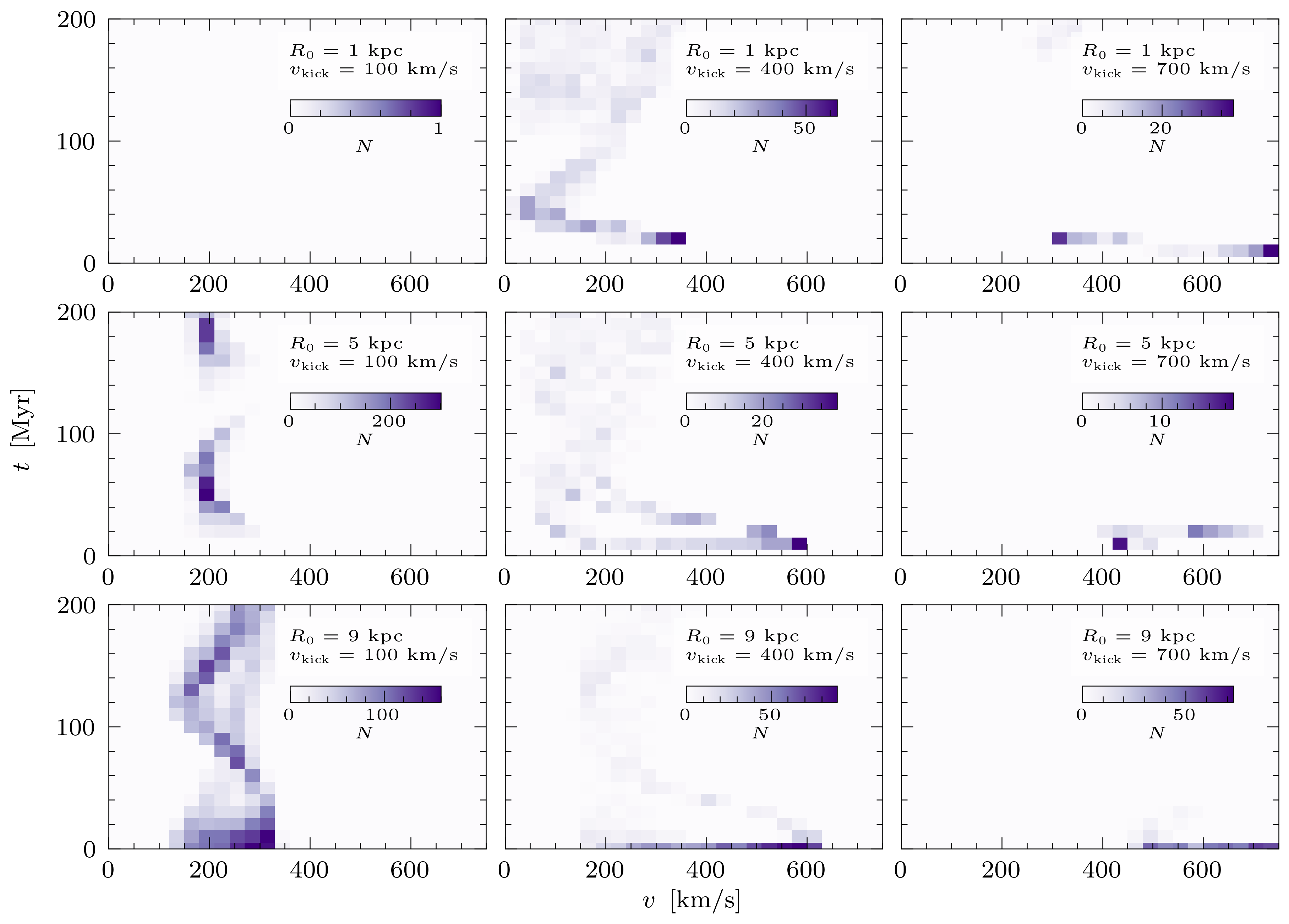}
    \caption{Grid of simulations of $10^3$ kicked objects, similar to Fig.\ \ref{fig6}, but showing only the objects which are in the Solar neighbourhood. Again, these simulations have a fixed initial Galactocentric radius $R_0$ (rows: $1$ kpc, $5$ kpc, and $9$ kpc) and a single value for $v_{\text{kick}}$ (columns: $100$ km/s, $400$ km/s, and $700$ km/s). Similarly to Fig.\ \ref{fig6}, the evolution of the Galactocentric velocity ($v$) is shown through 2D histograms with time-bins of $10$ Myr and velocity-bins of $30$ km/s. However, since the sizes of the populations in the panels differ significantly, they are shown on different colour scales.}
    \label{fig7}
\end{figure*}
\subsection{Velocity direction}
\label{sec3.2}
Besides the deceleration in velocity magnitude, we are also interested in how the cylindrical components of the velocity vectors (i.e., $v_{\phi}$, $v_{R}$, and $v_{z}$) change over time. In Fig.\ \ref{fig5}, we show the evolution of the velocity magnitude and the separate components. The top row in the figure shows how the velocity magnitude distributions evolve, and how they become relatively stable after $20$ to $30$ Myr. Also, in the total population the high-velocity tail is visible, as we also found in Fig.\ \ref{fig2}.\\
\indent The second row shows the azimuthal velocity ($v_{\phi}$), which at $t=0$ Myr is not symmetrically distributed around $v_{\phi}=0$ km/s: the added $v_{\text{circ}}$ term in Eq.\ \ref{eq2} shifts the distribution to positive $v_{\phi}$. After this initial shift, however, the azimuthal velocities decrease. In the total population, these velocities are lower than in the Solar neighbourhood or thick disc, because this population includes the point masses which escape the Galaxy and travel to large distances, meaning the centripetal force caused by the gravitational attraction of the Galaxy becomes relatively small.\\
\indent The third row shows the radial velocity ($v_{R}$). Interestingly, these distributions start symmetrically distributed around $v_{R}=0$ km/s, but start to increase to positive velocity, before decreasing again after approximately $20$ Myr. Here, the total population differs in the fact that the deceleration does not cause a significant fraction of the sample to get negative $v_{R}$, as opposed to the Solar neighbourhood and the thick disc. Again, this is due to the subset of point masses which get a large kick and escape (and therefore retain a positive $v_{R}$).\\
\indent The fourth, and final, row shows the vertical velocities ($v_{z}$). Similar to $v_{R}$, the vertical velocities start symmetrically distributed around $v_{z}=0$ km/s. However, in all three populations, both positive and negative $v_{z}$ decrease significantly, even within the first $10$ Myr. This creates a narrow peak around $v_{z}=0$ km/s.\\
\indent In general, we note the large similarities between the velocity distributions in all three populations, and argue that this implies the population of objects that get close to the Solar system is a relatively representative sample of the total population. Therefore, we argue that the observational bias, as proposed by \citet{Cordes_1986} and \citet{Lyne_1994}, is insufficient to explain the fact that older pulsars are observed to have lower velocities. Instead, the apparent deceleration seems to be caused by the actual deceleration of objects as their orbit is changed by the kick, after which they spend most of their time at larger Galactocentric radii, effectively exchanging kinetic energy for potential energy.
\section{Dynamics}
\label{sec4}
Our simulation shows several interesting aspects of the velocity evolution of kicked objects, such as the general deceleration and the initial increase in $v_{R}$. In order to provide explanations for these aspects, we build a grid of simulations for different $R_0$ and $v_{\text{kick}}$, so the effects of these parameters can be examined (Sect.\ \ref{sec4.1}). Also, we simulate objects which get a total initial velocity completely in one of the cylindrical directions, to show how the direction of the initial velocity affects the trajectory and velocity evolution (Sect.\ \ref{sec4.2}).
\subsection{Grid}
\label{sec4.1}
We are interested in how the different initial Galactocentric radii ($R_0$) and kick velocities influence the velocity evolution. In order to do this, we use a grid of smaller simulations ($10^3$ objects instead of $10^4$), for several values of $R_0$ (i.e., $1$ kpc, $5$ kpc, and $9$ kpc, instead of the initial positions shown in Fig.\ \ref{fig1}) and $v_{\text{kick}}$ (i.e., $100$ km/s, $400$ km/s, and $700$ km/s, instead of sampling from the \citet{Hobbs_2005} distribution). Figure \ref{fig6} shows the velocity evolution of the total population for each value of $R_0$ and $v_{\text{kick}}$. The initial distributions (i.e., at $t=0$ Myr) range from $|v_{\text{circ}}(R_0)-v_{\text{kick}}|$ to $v_{\text{circ}}(R_0)+v_{\text{kick}}$, because of the angle between the initial circular velocity at $R_0$ and the kick. For $t>0$ Myr, the objects can have velocities outside of this range.\\
\indent The figure shows that low kicks (i.e., $100$ km/s) perturb the initial circular orbits of the objects only slightly, because of which the objects start to oscillate around their initial orbit. However, larger kicks (i.e., $400$ km/s and $700$ km/s) perturb the initial orbits significantly, meaning the objects start to migrate through the Galaxy. That is, they will spend most of their time at Galactocentric radii larger than the initial $R_0$, causing the deceleration. The figure also shows that this deceleration is more rapid at lower $R_0$. After all, the velocities of objects born near the Galactic centre decrease more rapidly as they move outwards, due to the shape of the MW potential.\\
\indent Figure \ref{fig7} shows the Galactocentric velocities of the objects that are in the Solar neighbourhood, for the grid of simulations. For $v_\text{kick}=100$ km/s, objects can only enter the Solar neighbourhood if their perturbed orbit intersects with the region. When $v_{\text{kick}}=100$ km/s, this is not the case for initial radii, $R_0=1$ kpc, but for $R_0=5$ kpc the figure shows that part of the orbits do intersect with the Solar neighbourhood. This part is even larger if $R_0=9$ kpc, since these objects are born within the Solar neighbourhood. When $v_{\text{kick}}=400$ km/s, the objects that arrive in the Solar neighbourhood will have more eccentric orbits, and the figure shows the deceleration (similar to the deceleration shown in Figs.\ \ref{fig5} and \ref{fig6}). For $v_\text{kick}=700$ km/s, most objects get a velocity greater than the escape velocity of the MW, which means they temporarily cross the Solar neighbourhood, but most will not return to this region.\\
\indent These objects, i.e., the ones that receive a high kick, influence the high-velocity tail, and cause the main difference between the Solar neighbourhood population and the total population. Since the high-kicks only influence the high-velocity tail, and this tail does not contribute significantly to the population of objects observed in the Solar neighbourhood (as shown in Figs.\ \ref{fig2}, \ref{fig5}, and \ref{fig7}), the initial kick distribution can be changed greatly without altering the distribution of objects older than $20$ Myr, observed in the Solar neighbourhood (as we show in Appendix \ref{appB}).\\
\indent Moreover, Fig.\ \ref{fig7} shows how different initial conditions can bring objects to the Solar neighbourhood with identical Galactocentric velocity. For instance, both when $R_0=1$ kpc and $v_{\text{kick}}=400$ km/s, and when $R_0=9$ kpc and $v_{\text{kick}}=100$ km/s, objects with a velocity of approximately $200$ km/s can be found in the Solar neighbourhood. This shows that an observed velocity in the Solar neighbourhood of about $200$ km/s could both be the result of a $100$ km/s kick as well as a $400$ km/s kick. This means that, besides obvious statements such as the fact that an observed object of any velocity is extremely unlikely to be the result of a $100$ km/s kick at $R_0=1$ kpc, we find that in order to determine the initial conditions of an individual observed object (of unknown age) based on its velocity magnitude, one needs to assume (1) the distribution of initial positions, and (2) the distribution of kick velocities. In other words, one needs to weigh the simulations in our grid according to assumed distributions of $R_0$ and $v_{\text{kick}}$, before being able to determine the probable origin of an observed object with a certain velocity.The speed of an object alone contains little information about the kick at birth.
\subsection{Extrema}
\label{sec4.2}
\begin{figure*}
    \sidecaption
    \includegraphics[width=12cm]{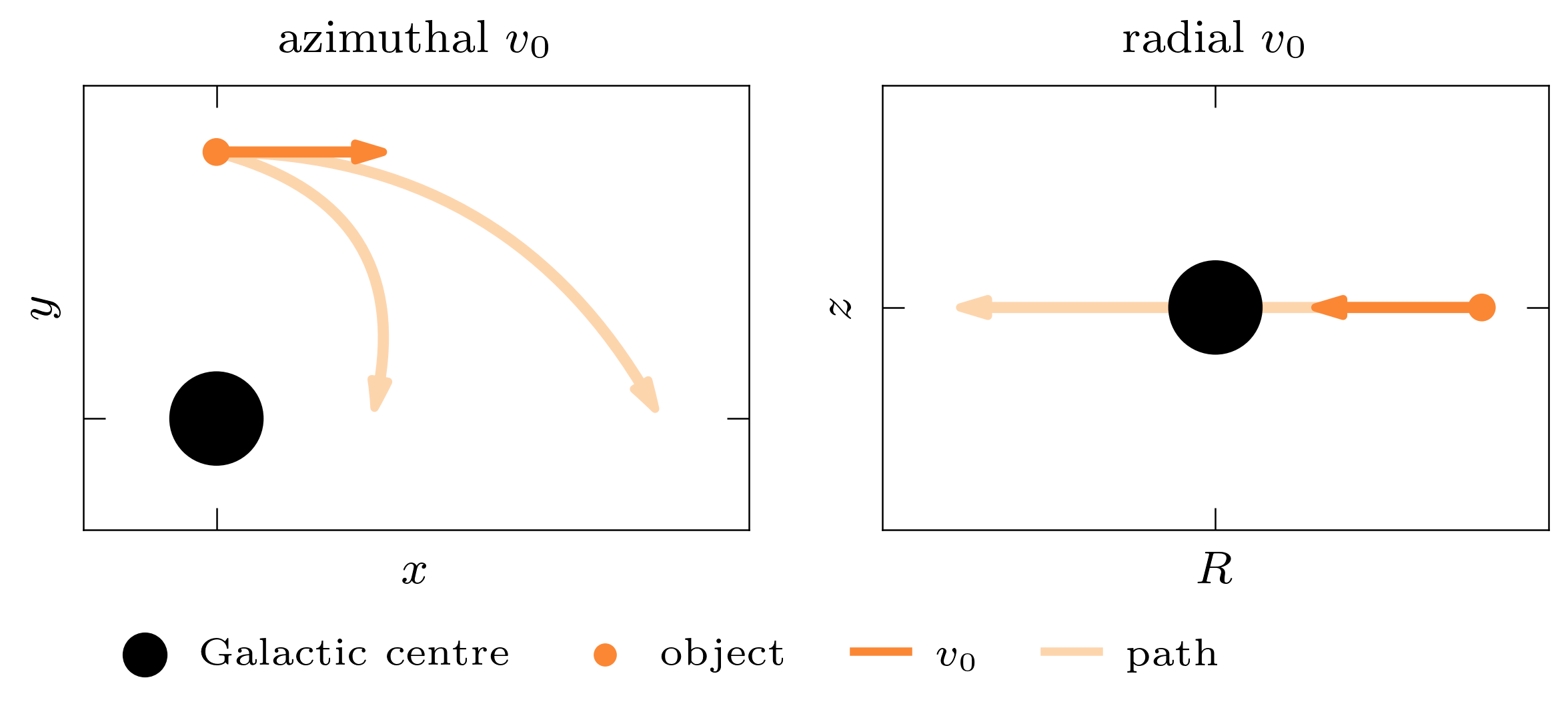}
    \caption{Schematic diagrams of hypothetical paths (light orange curves) of objects (orange dots), resulting from an initial velocity (orange arrows) completely in the azimuthal (left panel) or negative radial (central panel) direction. The shown initial velocity vectors and paths are merely qualitative estimates of possible paths that potentially contribute to the initial acceleration of $v_{R}$ (as found in Fig.\ \ref{fig5}). In Fig.\ \ref{fig9}, we show examples of orbits that are a result of initial velocities in these direction.}
    \label{fig8}
\end{figure*}
\begin{figure*}
    \centering
    \includegraphics[width=18cm]{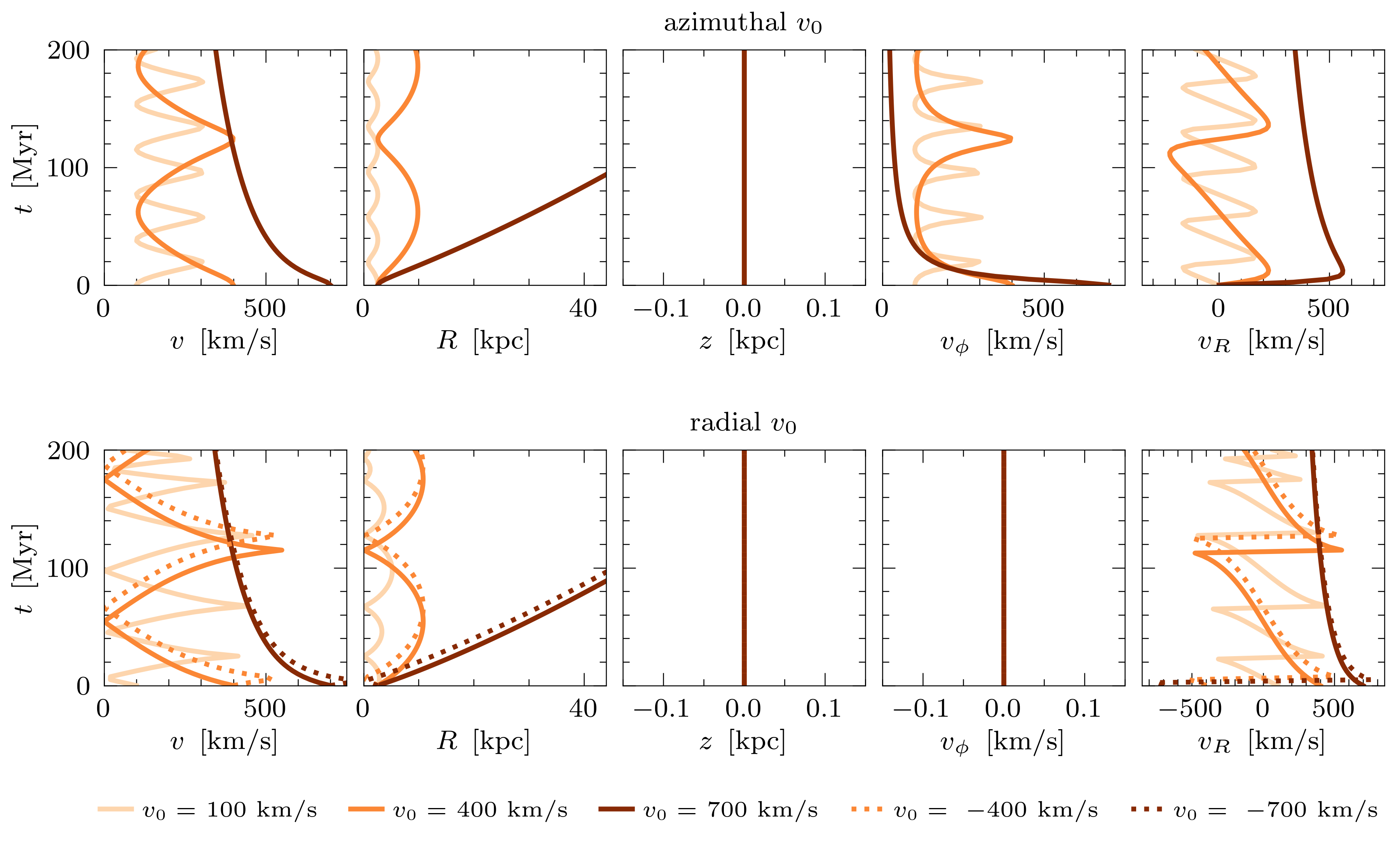}
    \caption{Trajectories of objects starting at $R_0=R_d=2.6$ kpc, which get an initial velocity ($v_0$) either completely in the $\phi$-direction (top row) or in the $R$-direction (bottom row). For each direction, we show the trajectories of $v_0=100$ km/s (light orange), $400$ km/s (orange), and $700$ km/s (dark orange), and add $-400$ km/s (dotted orange) and $-700$ km/s (dotted dark orange) for the radial direction. The first column shows the Galactocentric velocity ($v$), and the second and third column show the $R$ and $z$ positions of the objects, respectively. In the fourth column, we show the azimuthal velocity ($v_{\phi}$) and in the fifth column we show the radial velocity ($v_{R}$). The vertical velocity ($v_{z}$) does not obtain non-zero values in these orbits.}
    \label{fig9}
\end{figure*}
We are interested in explaining why the direction of the velocity vectors changes over time, as shown in Fig.\ \ref{fig5}. In particular, we are interested in explaining why initially $v_{R}$ increases, while $v_{\phi}$ and $v_{z}$ decrease immediately. In order to do this, we investigate the special cases where the initial velocity is pointed entirely in the $\phi$ or (negative) $R$ directions. In other words, in these two dimensions we look at the extrema of the orientation of the initial velocity vector (i.e., initial velocities, being a combination of $v_{\text{kick}}$ and $v_{\text{circ}}$ as defined in Eq.\ \ref{eq2}, pointed entirely in one of the cylindrical directions).\\
\indent We hypothesise that the reason $v_{R}$ initially increases while $v_{\phi}$ and $v_{z}$ decrease, is the fact that the orientation of the velocity vector of a kicked object changes within the cylindrical coordinate system, because of their motion through the Galactic potential. The cylindrical coordinate system can, then, favour reorientation in the positive $R$-direction, causing the acceleration. In Fig.\ \ref{fig8}, we sketch how initial velocity vectors oriented in the $\phi$ and negative $R$ direction can be reoriented by their motion through the Galactic potential in such a way that they obtain a positive $v_{R}$ component.\\
\indent The figure shows the following two effects which can reorient the velocity vector in the positive $R$ direction (as sketched in the panels of Fig.\ \ref{fig8}):
\begin{itemize}
    \item If the initial velocity, $\vec{v}_0$, is oriented in the azimuthal direction, the probability of the velocity being exactly the one needed for a circular orbit is extremely small. This means that the resulting orbit will not be circular: if $v_0>v_{\text{circ}}$, the object will travel to larger $R$, and if $v_0<v_{\text{circ}}$, the object will fall inwards to smaller $R$. The initially azimuthal velocity will therefore probably reorient in the cylindrical coordinate system, obtain a radial component, and, because a \citet{Hobbs_2005} kick is typically larger than $v_{\text{circ}}$, this radial component will most likely be positive.
    \item If $\vec{v}_0$ is oriented in the negative radial direction, it will travel towards the Galactic centre and overshoot to the positive $R$ direction (neglecting possible collisions with other objects). Because of the cylindrical coordinate system, negative radial velocity will turn into positive radial velocity, while this is not true vice versa.
\end{itemize}
Because of these effects, the vectors of azimuthal and negative radial velocities are able to obtain a positive radial component. However, positive radial motion does not reorient itself in a similar way. Positive radial motion can either escape the Galaxy, or decelerate enough to turn into negative radial motion.\\
\indent In order to reproduce these effects in our simulation, we examine the orbits of objects that get an initial velocity entirely in one of these cylindrical dimensions. We put objects at $R_0=2.6$ kpc--since this is the half-light radius of the disc, approximating the peak in Fig.\ \ref{fig4}--and give them initial velocities of $100$ km/s, $400$ km/s, and $700$ km/s, in the azimuthal or radial direction. Figure\ \ref{fig9} shows the orbits resulting from these initial conditions.\\
\indent The figure shows that azimuthal velocity can indeed reorient into radial velocity. Within the first $20$ Myr, the azimuthal velocity can decrease rapidly while the radial velocity increases, and this effect is greatest for the largest kick. The smallest $v_0$ (i.e., $100$ km/s) is smaller than the local $v_{\text{circ}}$, causing the object to fall inward (as sketched in Fig.\ \ref{fig8}). The radial velocity, in contrast, can not reorient into other dimensions. However, as the simulated orbits show, negative radial velocity can turn positive when crossing the Galactic centre, and positive radial velocity can turn negative when it is decelerated completely and falls back to the Galactic centre. As Fig.\ \ref{fig9} shows, the former tends to happen on a shorter timescale, causing an increase in $v_{R}$ before $20$ Myr. Then, these objects are decelerated, where the object with the highest $v_0$ (i.e., $700$ km/s) escape the Galaxy.\\
\indent Kicked objects can of course also obtain a $v_0$ in the $z$ direction. Because of the gravitational attraction to the Galactic centre, these orbits are also able to obtain a positive $v_R$ component. However, for most kicks this will mostly occur at positions with a relatively large $|z|$, i.e., outside of the Solar neighbourhood. This effect, therefore, probably does not contribute significantly to the initial increase in $v_R$, since this increase looks similar in the Solar neighbourhood and the total population.\\
\indent Moreover, if the kick vectors are oriented isotropically, the probability of the initial velocity being pointed entirely in one of the directions is negligible. We therefore do not argue that the discussed effects are a comprehensive description of the simulated orbits, but instead conclude they do show that the velocity vector initially tends to reorient in the positive radial direction, which can help explain the initial increase in $v_R$ in our simulation.
\begin{figure*}
    \centering
    \includegraphics[width=18cm]{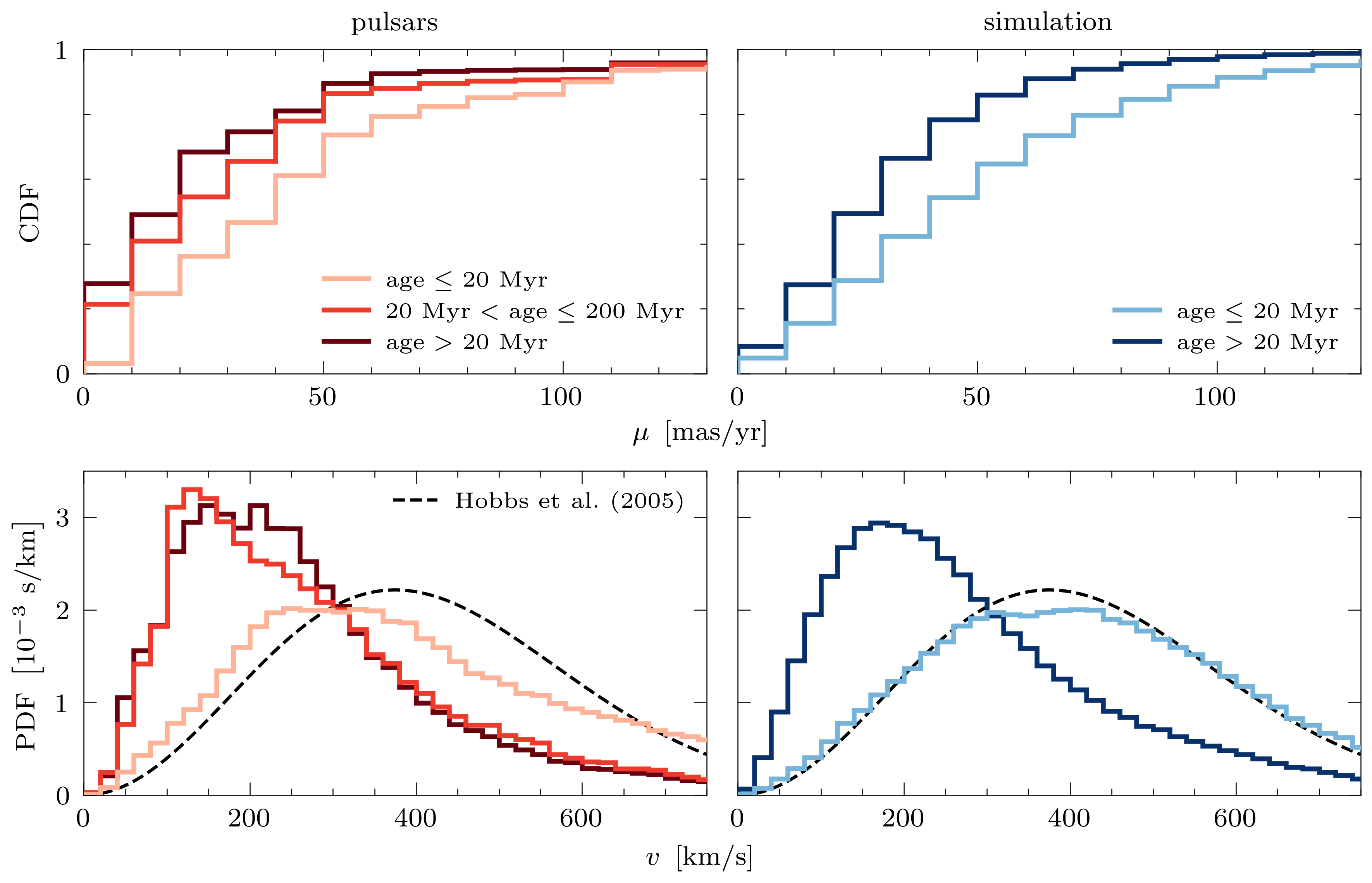}
    \caption{Comparison between pulsar data and our simulation. The pulsar data is taken from the ATNF catalogue \citep{Manchester_2005}, as described in the text. The top left panel shows the observed proper motions ($\mu$) of pulsars within $2$ kpc of the Solar system, for ages below $20$ Myr (light red, consists of $72$ pulsars), between $20$ Myr and $200$ Myr (red, consists of $22$ pulsars), and above $20$ Myr (dark red, consists of $49$ pulsars), in cumulative histograms with a bin-width of $10$ mas/yr. In the top right panel, we show the proper motions of the simulated objects in the Solar neighbourhood, for ages below $20$ Myr (light blue) and above $20$ Myr (dark blue), in a similar histogram. The bottom left panel shows the sum of the Monte Carlo estimated posteriors of the pulsar Galactocentric velocities ($v$), in histograms with a bin-width of $20$ km/s, together with the \citet{Hobbs_2005} kick distribution. These distributions also include pulsars at distances greater than $2$ kpc from the Solar system, resulting in populations with ages below $20$ Myr, between $20$ Myr and $200$ Myr, and above $200$ Myr with sizes of $179$, $45$, and $101$, respectively. The bottom right panel shows similar histograms, for the total population in our simulation, as is also shown in Fig.\ \ref{fig2}. Similarly to Fig.\ \ref{fig2}, we evaluate our simulation every $1$ Myr for the first $20$ Myr, after which we evaluate the simulation every $10$ Myr.} 
    \label{fig10}
\end{figure*}
\section{Pulsars}
\label{sec5}
Since we have found that, in our simulation, kicked objects decelerate due to the Galactic potential, we can investigate if any such signal is visible in observational data. In order to do this, we take a look at pulsar data, since they experience kicks and their age can be estimated. We convert the 3D velocities from our simulation into 2D proper motions, which we compare to observed proper motions of pulsars (Sect.\ \ref{sec5.1}). Then, we use a Monte Carlo estimation to determine the 3D Galactocentric velocities of the pulsars, and compare these to our simulation (Sect.\ \ref{sec5.2}). 
\subsection{Proper motions}
\label{sec5.1}
We use the pulsar data from the ATNF pulsar catalogue \citep{Manchester_2005}, and select all the pulsars with estimates for position, proper motion, distance, and age, except those with a binary companion. In total, the ATNF pulsar catalogue contains $280$ pulsars which meet these requirements. We compare 2D proper motions of the Solar neighbourhood population in our simulation with the pulsars that have estimated distances below $2$ kpc, and compare the total population in our simulation with estimated 3D velocities of all $280$ pulsars.\\
\indent In order to convert the Galactocentric 3D velocities of our simulation into 2D proper motions, we use \lstinline{ASTROPY}\footnote{\href{http://www.astropy.org}{http://www.astropy.org}} \lstinline{v.6.0.0} \citep{Astropy_2013,Astropy_2018,Astropy_2022} to convert the simulated objects in the Solar neighbourhood from a Galactocentric frame to a heliocentric frame, converting the 3D velocity vectors into 2D proper motion vectors. Since we need to account for all $12$ spheres that we use to determine the objects in the Solar neighbourhood, we rotate the coordinates in our simulation by $n\cdot2\pi/12$ for each $n^{\text{th}}$ sphere and add the proper motions of the objects within $2$ kpc of the Solar system to the distribution. We also correct for the motion of the sun, by rotating the objects at time $t$ by $-t\cdot v_{\phi,\sun}/R_{\sun}$ (using the values from Eq.\ \ref{eq3} and Eq.\ \ref{eq4}).\\
\indent For each pulsar, we sample $10^3$ values from Gaussians describing the observed proper motion and its uncertainty. After doing this for the proper motion in both right ascension ($\mu_{\alpha}$) and declination ($\mu_{\delta}$), and combine them to get the total 2D proper motion vector $\left(\mu^2=\mu_{\alpha}^2+\mu_{\delta}^2\right)$. Then, we sum all posterior distributions of pulsars with a certain age to create the proper motion distributions of young and old pulsars.\\
\indent There are many reasons why the pulsar data should not match our simulation. For example:
\begin{itemize}
    \item In our simulation, we assume the kick velocities are distributed according to the \citet{Hobbs_2005} distribution. However, it is still being debated whether this distribution accurately describes neutron star kicks, with more recent studies finding alternative distributions, some of which are bimodal \citep[e.g.,][]{Verbunt_2017,Igoshev_2021,O'Doherty_2023}.
    \item While our simulation has cylindrical symmetry, with the Galactic centre in the origin, the pulsar data has a bias towards nearby pulsars. After all, the closer a pulsar is to the Solar system, the lower the luminosity threshold for detectability \citep[e.g.,][]{Chrimes_2021}.  
    \item The distance estimates used in the ATNF catalogue may be inaccurate \citep[as e.g.\ argued by][]{Verbunt_2017}, which would influence our estimation of the 3D velocity based on the 2D proper motion.
    \item The pulsar ages stated in the ATNF catalogue \citep{Manchester_2005} are the characteristic spin-down ages, which equal $P/(2\dot{P})$ for a period $P$, and assume that the braking index equals $3$, and the initial period is significantly smaller than the current period \citep[e.g.,][]{Ostriker_1969,Shapiro_1983}. However, these assumptions may not be completely accurate for all pulsars \citep[e.g.,][]{HESS_2011}, meaning the true age of a pulsar probably differs from the characteristic age \citep{Jiang_2013}.
    \item The Monte Carlo estimates of the 3D pulsar velocities (as described in Sect.\ \ref{sec5.2}) assume an isotropically distributed velocity vector. However, our simulation shows that even after a few Myr, the velocity vectors are far from isotropic (e.g., Fig.\ \ref{fig5}).
    \item There may be a spin-kick alignment \citep{Johnston_2005}, which means that there is an observational bias towards pulsars moving radially with respect to us. Furthermore, if kicked objects preferentially end up on more radial orbits this contradicts the assumption of isotropy, causing an underestimation of the kick velocity \citep{Mandel_2023}. 
\end{itemize}
However, despite all of these arguments for differences, we find that the pulsar data and our simulation match reasonably well.\\
\indent Figure\ \ref{fig10} shows the cumulative distributions of the proper motions, and comparing the observed pulsars younger than $20$ Myr to the ones older than $20$ Myr shows a deceleration similar to the one in our simulation. The proper motion distribution of points in our simulation below $20$ Myr approximates the distribution of pulsars younger than $20$ Myr. For older pulsars, there is only a small difference between the distribution of pulsars with ages in between $20$ Myr and $200$ Myr, and the distribution of all pulsars older than $20$ Myr (which also includes pulsars older than $200$ Myr); both of these distributions are well-described by our simulation. This means that, despite all the observational biases and other reasons for differences between simulation and observation, they match relatively well. 
\subsection{Galactocentric velocities}
\label{sec5.2}
Besides the 2D proper motion comparison, we also investigate whether there is an agreement between the 3D Galactocentric velocities of the pulsars and our simulation. However, only the 2D proper motion can be observed, meaning we only know the projection of the 3D velocity vector on the sky. In order to estimate the Galactocentric velocity based on the proper motion, we use a Monte Carlo estimation. For each pulsar, we construct three Gaussians that correspond to the mean and uncertainty of the proper motion ($\mu_{\alpha}$ and $\mu_{\delta}$), and distance to the pulsar ($D$), as stated in the ATNF catalogue. Then, we sample $10^3$ values from these Gaussian distributions (where we require that $D>0$), based on which we determine the transverse part of the 3D velocity vector. In order to determine the total 3D velocity, we assume Galactocentric isotropy and uniformly sample a value $u$ between $-1$ and $1$, for each set of values of $\mu_{\alpha}$, $\mu_{\delta}$, and $D$, after correcting for the motion of the Solar system we set the angle between the transverse velocity and the total velocity to be $\cos^{-1}(u)$ \citep[see][for a more detailed description]{Gaspari_2024}. We determine the magnitude of the 3D Galactocentric velocity vector for all $10^3$ samples, resulting in a posterior velocity distribution for each pulsar. For comparison between the observations and our simulation, we sum the posteriors of pulsars with a certain age.\\
\indent Figure\ \ref{fig10} shows the results of our Monte Carlo estimation, including the posteriors for pulsars with ages below $20$ Myr, in between $20$ Myr and $200$ Myr, and above $20$ Myr (including pulsars older than $200$ Myr). The young pulsars start with a velocity distribution relatively close to the \citet{Hobbs_2005} kick distribution, though not as close as our simulation. Moreover, the older pulsars show a decelerated velocity distribution, which looks similar to the decelerated distribution of our simulation (for the total population). Both the posteriors of the observations and the distribution in our simulation show a peak around $200$ km/s.\\
\indent The fact that our simulation resembles the pulsar data well (especially for older pulsars), despite all the reasons why it should not, is either coincidental or implies that the deceleration due to the Galactic potential is a major factor in shaping the velocity distributions of pulsars older than $20$ Myr. Some aspects of the match between simulation and observation are accidental. For example, the high-velocity tail in the total population of our simulation consists of objects which get a high kick and escape the Galaxy, while the high-velocity tail in the pulsar data is more likely due to uncertainty in our Monte Carlo estimation. Nevertheless, we argue that the general match between the deceleration seen in the pulsars and the deceleration in our simulation, both in proper motion and Galactocentric velocity, is unlikely to be coincidental, and shows how deceleration by the Galactic potential has a major influence on the pulsar velocities. After all, we find that the pulsar velocities are well described by the velocity distribution of the total population, meaning observational biases or selection effects do not influence the results significantly.\\
\indent It is not surprising that the distributions of older pulsars match our simulation better than the distribution of young pulsars, since, as we mentioned in Sects.\ \ref{sec3.1} and \ref{sec4.1}, we find that the velocity distribution of older objects in the Solar neighbourhood remains constant when varying the kick distribution (Appendix \ref{appB}). In other words, the velocities of older pulsar may not depend significantly on the kick distribution of \citet{Hobbs_2005} being accurate.
\section{Conclusion}
\label{sec6}
In this work, we simulated the trajectories of objects that migrated through the Galaxy after receiving a velocity kick (Sect.\ \ref{sec2}) and found that they are decelerated by the Galactic potential (Sect.\ \ref{sec3}). This deceleration is a result of the kicks which increase the trajectories' apocentres (Sect.\ \ref{sec4}), and is consistent with the deceleration observed in the velocities of Galactic pulsars at later ages (Sect.\ \ref{sec5}). Based on this, we conclude the following:
\begin{itemize}
    \item Objects that receive a kick tend to have their trajectories' apocentres increased, when they do not get unbound. As a result, they spend most of their time at larger radii, where they have lower velocities and are hence decelerated with respect to their initial velocity (Fig.\ \ref{fig3}). This causes the velocity distribution to shift toward lower values within 20 to 30 Myr from the kick, after which the median velocity stabilises at around $200$ km/s (Fig.\ \ref{fig2}).
    \item In the Solar neighbourhood, objects older than about $20$ Myr are, approximately, a representative sample of the total population, both in terms of velocity (Fig.\ \ref{fig2}) and initial Galactocentric radius (Fig.\ \ref{fig4}).
    \item Objects that receive a small kick (e.g., 100 km/s) are only slightly disturbed in their initial circular orbit, while higher kicks (e.g., 400 km/s or higher) disturb the initial orbit significantly, and result in the deceleration described above. Therefore, objects that received small kicks keep travelling at around the same initial speed, while the slowest objects only result from the upper-tail of the kick distribution (Fig. \ref{fig7}).
    \item Within the first $20$ Myr, $v_{\phi}$ and $v_{z}$ decrease while $v_{R}$ increases (Fig.\ \ref{fig5}). This is due to the reorientation of the initial velocity vector into the positive radial direction (Fig.\ \ref{fig8}), and suggests that nearby objects have non-isotropic velocities within this age.
    \item Objects in the Solar neighbourhood with identical velocity and age can have a vastly different origin, in terms of $v_{\text{kick}}$ and $R_0$ (Fig.\ \ref{fig7}).
    \item Using the Galactic pulsars from the ATNF catalogue \citep{Manchester_2005}, we find that both the observed 2D proper motions as well as the estimated 3D Galactocentric velocities of the pulsars resemble our simulation surprisingly well (Fig.\ \ref{fig10}). This indicates that the deceleration due to migration through the Galactic potential plays a major role in determining the pulsar velocities.
\end{itemize}
As a more general conclusion, we argue that the observational bias posed by \citet{Cordes_1986} and \citet{Lyne_1994} --i.e., the hypothesis that high-velocity objects leaving the Solar neighbourhood sooner causes a selection effect-- is insufficient in explaining the apparent deceleration of pulsars as they age. Instead, we find that the difference between the Solar neighbourhood and the total population of kicked objects is merely the high-velocity tail, and the deceleration can be found in all populations: producing a peak at approximately $200$ km/s. Therefore, we argue that the deceleration of pulsars as they age is not a selection effect but a genuine deceleration, caused by the Galactic potential.\\
\indent For future research, it can be interesting to further investigate the problem of how to determine the kick velocity of an object older than $20$ Myr, based on its observed proper motion and position. After all, we find that the (scalar) velocity distribution of objects in the Solar neighbourhood is mostly insensitive to the kick distribution (Appendix \ref{appB}). Since our simulation is a relatively simplistic one, it can be compared to any population of kicked objects, such as binaries and--as we discuss in this work--pulsars. Such a comparison could potentially help explain the observed velocities of these binaries \citep[e.g.,][]{Zhao_2023} and pulsars \citep[e.g.,][]{Cordes_1998}. Moreover, it could be interesting to repeat the Monte Carlo estimation of the pulsar velocities, but use the distributions of $v_{\phi}$, $v_{R}$, and $v_{z}$ we show in Fig.\ \ref{fig5}, instead of assuming isotropy.

\begin{acknowledgements}
    We thank Frank Verbunt, Gijs Nelemans, and Ilya Mandel for useful discussions regarding this Master's project, and the referee, Andrei Igoshev, for valuable comments that helped improve this paper. AJL was supported by the European Research Council (ERC) under the European Union’s Horizon 2020 research and innovation programme (grant agreement No. 725246), and NG acknowledges studentship support from the Dutch Research Council (NWO) under the project number 680.92.18.02. In this work, we used pulsar data from the ATNF catalogue \citep[][available at \href{https://www.atnf.csiro.au/research/pulsar/psrcat}{https://www.atnf.csiro.au/research/pulsar/psrcat}]{Manchester_2005}. Moreover, we made use of \lstinline{NUMPY} \citep{Harris_2020}, \lstinline{MATPLOTLIB} \citep{Hunter_2007}, \lstinline{GALPY} \citep{Bovy_2015}, and \lstinline{ASTROPY}, a community-developed core Python package and an ecosystem of tools and resources for astronomy \citep{Astropy_2013,Astropy_2018,Astropy_2022}.
\end{acknowledgements}
\balance
\bibliographystyle{TeXnical/aa_url}
\bibliography{References}

\begin{appendix}
\nobalance
\section{Velocity evolution}
\label{appA}
\begin{figure}[h]
    \resizebox{\hsize}{!}{\includegraphics{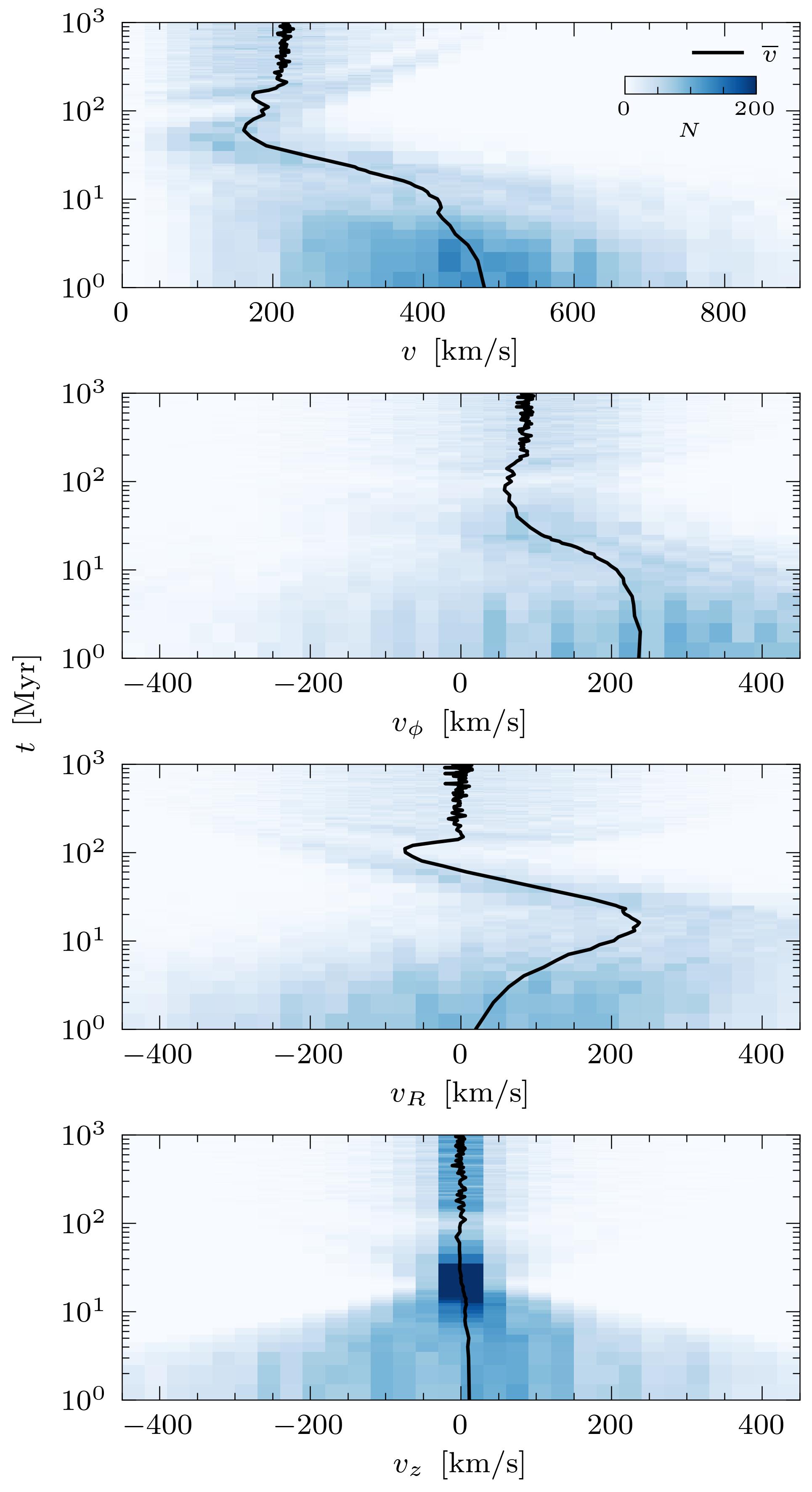}}
    \caption{Velocity evolution of the objects in the Solar neighbourhood, similar to Fig.\ \ref{fig5}, but on a logarithmic scale extended to $1$ Gyr. The panels show the magnitude of the velocity vector ($v$), as well as the cylindrical velocity components ($v_{\phi}$, $v_{R}$, and $v_{z}$). The first $25$ Myr, the velocities are evaluated every $1$ Myr, after which they are evaluated every $10$ Myr, as shown in the 2D histogram with velocity-bins of $30$ km/s. The black line traces the mean velocity ($\overline{v}$) at each point in time.}
    \label{figA}
\end{figure}

In order to examine the deceleration in the velocity distribution more closely, we show the velocities of objects in the Solar neighbourhood in Fig.\ \ref{figA}, similarly to Fig.\ \ref{fig5} but on a logarithmic axis which is extended to $1$ Gyr. The figure shows how the mean of the scalar velocity starts to decrease rapidly after about $10$ Myr, and stays at approximately $200$ km/s. This shows how, despite the fact that the system may not have reached full dynamical equilibrium yet \citep[e.g.,][ch.\ 4]{Binney_1987}, the deceleration to mean velocities of $\sim200$ km/s in the solar neighbourhood appears to be stable beyond $200$ Myr. This is true for the evolution in the cylindrical velocity components as well.
\section{Sensitivity to initial kick}
\label{appB}
\begin{figure*}
    \centering
    \includegraphics[width=18cm]{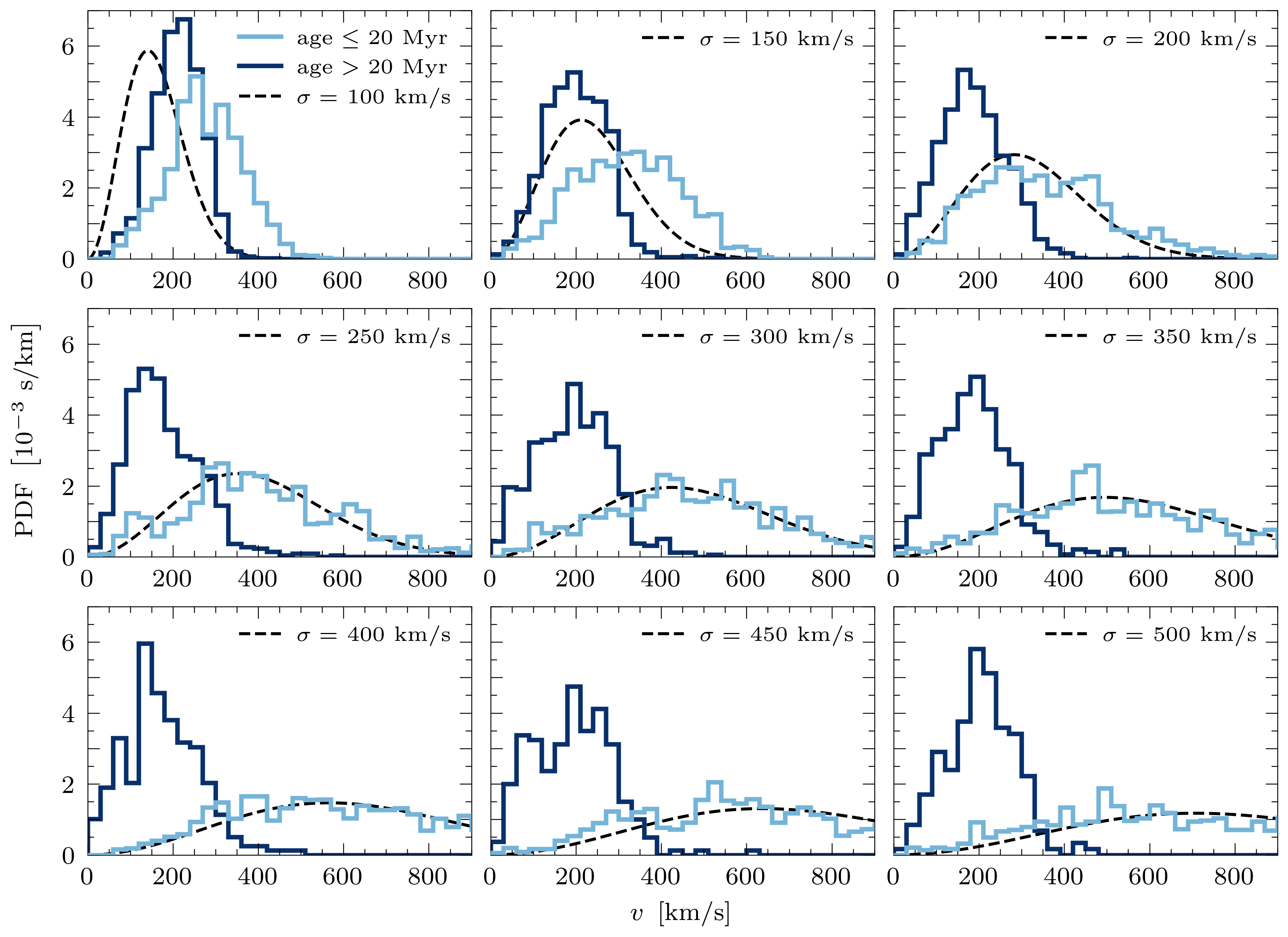}
    \caption{Our simulation, but for $10^3$ point masses instead of $10^4$, and for varying kick distributions. Similarly to the results shown in Fig.\ \ref{fig2}, we evaluate the points below $20$ Myr every $1$ Myr (light blue curve), and after $20$ Myr every $10$ Myr, which provides the velocity distributions shown in normalised histograms with a bin-width of $30$ km/s. For the kick distributions, we use Maxwellians (Eq.\ \ref{eqB1}) with a $\sigma$ varying from $100$ km/s to $500$ km/s, with intervals of $50$ km/s.}
    \label{figB}
\end{figure*}
In our simulation, we use the \citet{Hobbs_2005} kick distribution, which is a Maxwellian with a $\sigma$ of $265$ km/s. However, we find that the peak at approximately $200$ km/s (as shown in Fig.\ \ref{fig2}) also forms if we change the kick distribution. In order to show this, we run our simulation for different kick distributions, all described by Maxwellians: 
\begin{equation}
    \label{eqB1}
    f(v)=\sqrt{\dfrac{2}{\pi}}\dfrac{v^2}{\sigma^3}\exp\left(-\dfrac{v^2}{2\sigma^2}\right)\quad,
\end{equation}
with values for $\sigma$ varying from $100$ km/s to $500$ km/s. Figure\ \ref{figB} shows the results of varying the kick distribution in this way for the observed velocity distributions in the Solar neighbourhood. As can be seen in the figure, varying the kick distribution does not affect the decelerated distribution meaningfully, since the distributions above $20$ Myr are all nearly indistinguishable. This is not surprising, since a larger $\sigma$ increases the fraction of the total population which get a high enough $v_{\text{kick}}$ to escape the Galaxy, and after all, these are not observed in the Solar neighbourhood. We find, therefore, that it is extremely difficult to estimate the kick distribution based on the magnitude of the velocities of observed objects in the Solar neighbourhood that are older than $20$ Myr, although it might be possible to constrain kicks based on the complete velocity vector \citep[e.g.,][]{Atri_2019}.
\end{appendix}

\end{document}